\documentclass[aip,preprint,groupedaddress]{revtex4}
\usepackage{amssymb}
\usepackage{amsmath}
\usepackage{graphicx}
\usepackage{graphics}
\usepackage{epsfig}
	
\begin{document}

\title{Simulation of electronic and geometric degrees of freedom using a kink-based path integral formulation: application to molecular systems}


\author{Randall W. Hall}
\affiliation{Department of Chemistry\\Louisiana State
University\\Baton Rouge, La 70803-1804}


\date{\today}

\begin{abstract}
A kink-based path integral method, previously applied to atomic systems, is modified and used to study molecular systems.  The method allows the simultaneous evolution of atomic and electronic degrees of freedom.  Results for CH$_4 $, NH$_3 $, and H$_2 $O demonstrate this method to be accurate for both geometries and energies.  Comparison with DFT and MP2 level calculations show the path integral approach to produce energies in close agreement with MP2 energies and geometries in close agreement with both DFT and MP2 results.
\end{abstract}


\maketitle


\section{Introduction}
The development of simulation methods that are capable of treating electronic degrees of freedom at  finite temperatures is necessary for the study of a variety of important systems including those with multiple isomers with similar energies (such as metal clusters) and with dynamic bond breaking/forming processes.  A fundamental difficulty in using \textit{ab initio} quantum
approaches to study systems at finite temperatures is the need for
most algorithms to solve a quantum problem (to find, for example, the
\textit{ab initio} forces) at each geometric
configuration. Thus the CPU requirement  per time or Monte Carlo step often 
prevents a simulation.  Feynman's path integral formulation of quantum mechanics\cite{fandh} offers the
possibility of simultaneously treating geometric and electronic degrees of freedom without the restriction of solving a quantum problem for fixed atomic positions.  In addition, temperature and
electron-electron correlation can be included  and make this
approach very tempting as a starting point for \textit{ab initio}
simulations. An unfortunate aspect of the path integral approach is
the so-called ``sign problem'' which can make the standard deviations
of estimated quantities (such as energy) too large for practical
use\cite{piMak1,piMak2,piMak3,piOkazaki,ceperleypiH,kukipi,rwhpihub,rwhpiks,ceperley3he,voth1,voth2,ceperleyh,ceperleyfixednode,AFMC1,AFMC2,AFMC3,AFMC4,pikink1,pikink2,Deymier2004,Neuhauser2000,Baer2004}.
This problem occurs because the quantum density matrix is not
positive definite and results in averages being determined from
sums of large numbers with different signs.  The Car-Parinello
 implementation of density functional theory\cite{CPREV} is motivated by needs similar to the ones described above  and  treats 
electronic and geometric degrees of freedom on a similar footing and allows both types of degrees of freedom to
propagate simultaneously during a calculation.  A limitation of this
approach, though, is the need for the electronic degrees of freedom,
as described using single particle orbitals, to
be very close to the lowest energy set of orbitals, forcing the use of
small time steps in a molecular dynamics simulation.

We have recently introduced\cite{pikink1,pikink2} a "kink-based" path integral approach 
and have demonstrated that it can be used  to overcome the "sign problem" in atomic
systems. In the present work  a
formalism appropriate for  molecular systems is developed. To construct a
practical approach, an approximation to the exact path
integral approach is made; the approximation is based on the results of our
previous work.  The method treats the electronic structure as consisting of ground and excited single determinant states built from atom-based orbitals.  Simulations include moves that perform unitary transformations of the single particle orbitals, additions and deletions from a list of excited states used to evaluate the energy, and moves of atoms.   Using this procedure, electronic and geometric degrees of freedom are treated simultaneously.  The method is used with success to calculate the average energies and geometries of  CH$_{4}$, NH$_3 $, and H$_2 $O at finite temperatures.  We define
success as (a) overcoming the sign problem, (b) not requiring low
energy orbitals, (c) average molecular geometries in agreement with
previous \textit{ab initio} calculations, and
(d) average energies that compare favorably with  previous
\textit{ab initio} calculations.

\section{Kink-based Path Integral Formulation}
Our previous work\cite{pikink1,pikink2} started with the path integral expression for the  canonical
partition function, evaluated with fixed geometries and
in a discrete N-electron basis set$\{|j>\}$:
\begin{eqnarray}
Q(\{|j>\})&=&Tr\left\{\exp(-\beta H)\right\}=\sum_j <j|\exp(-\beta H)|j>
\end{eqnarray}
Making the Trotter approximation and discretizing the path into $P$ segments, we find
\begin{eqnarray}
Q\left( P,\{|j>\}\right)  &=& \left(\prod_{i=1}^{P} \sum_{j_i }\right)<j_i |\exp(-\beta H/P)|j_{i+1}>  
\end{eqnarray}
This can be interpreted as a path in the space of states that starts and ends with state $|j_{1}>$. If $|j_i > = |j_{i+1}>$, we have a diagonal matrix element, otherwise we have an off-diagonal matrix element. Any place that an off-diagonal element appears is called  a "kink" and it is clear that the paths can be classified into those paths with zero kinks, 2 kinks, 3 kinks, etc.  We demonstrated\cite{pikink1,pikink2} that in terms of kinks  
\begin{eqnarray}
Q\left( P,\{|j>\}\right)  &=&\sum_{j}x_{j}^{P}+  \nonumber \\
&&\sum_{n=2}^{P}\frac{P}{n}\left(
\prod_{i=1}^{n}\sum_{j_{i}}\right) \left(
\prod_{k=1}^{n}t_{j_{k},j_{k+1}}\right)\nonumber\\
&\times&\prod_{k=1}^{m}\left[ \frac{1}{\left( s_{j_{k}}-1\right) !}\frac{%
d^{s_{j_{k}}-1}}{dx_{j_{k}}^{s_{j_{k}}-1}}x_{j_{k}}^{s_{j_{k}}-1}\right]
\sum_{l=1}^{m}\frac{x_{j_{l}}^{P-n+m-1}}{\prod\limits_{i\neq
l}\left( x_{j_{l}}-x_{j_{i}}\right) }
\label{originalQ}
\end{eqnarray} with
\begin{eqnarray}
x_j &=&<j|\exp(-\beta H/P)|j>\\
t_{ij}&=&<i|\exp(-\beta H/P)|j>
\end{eqnarray}
$\beta = 1/k_{B}T$, $P$ is  the discretizing variable in the path
integral formulation,  $n$
is the number of kinks, $s_j $ is the number of times a particular
state appears in the sum, and $m$  is the number of distinct states
that appear ($\sum_{j=1}^m s_j = n$).  As written,
Eqn.~\ref{originalQ} is amenable to a Monte Carlo simulation.
However, since the matrix elements can be negative, the usual sign
problem will occur if the states are not well chosen. In our
previous work\cite{pikink1,pikink2}, the initial N-electron states $|j>$ were chosen
to be simple, anti-symmetrized products of 1-electron orbitals. The 
 N-electron states were improved by
periodically diagonalizing the Hamiltonian in the space of those
states that occurred during the simulation.  The
result was that the final states were linear
combinations of the initial orbitals (essentially the "natural spin orbitals" for the system), the only paths that appeared
at the end of the simulation contained 0, 2, or 3 kinks, and the
sign problem was reduced to insignificance. Further, the
dimensions of the density matrix were small enough that the matrix
could be kept in memory and transformed whenever a diagonalization
took place; this did not require any matrix elements involving the
initial orbitals, which significantly reduced the computational
effort.

In the case of geometric degrees of freedom, all matrix elements
are referenced to the initial orbitals and the adaptive scheme
used for atomic systems must be modified for computational
efficacy. Our previous work showed that once a good guess for the ground state was obtained 
the vast majority of paths contained 0 or 2 kinks. Using an approximate Hartree-Fock solution as a guess for the ground state, an
approximate infinite order summation of kinks from the ground
state is developed (leaving for future work the straightforward extension to the case of degenerate or nearly degenerate
ground states), in which we assume the most important  paths contain many instances of the ground state.

First consider a Hartree-Fock-like approximation to $Q(P)$.
In the Born-Oppenheimer approximation, $Q$ is given by
\begin{eqnarray}
Q &=&Tr\left\{\exp\left(-\beta H\right)\right\}=\int\mathbf{dR}^N \sum_{j }<j | \exp\left(-\beta H(\mathbf{R}^N \right)|j >
\end{eqnarray}
where we assume N nuclei, N$_e $ electrons, and that $\{|j >\}$ is a set of
N$_e $-electron orbitals. Each N$_e $-electron orbital is expressed as an
anti-symmetrized product of 1-electron orthonormalized spin-orbitals
$|j>=A(\phi_{j_{1}}\alpha_{j_{1}}  , \phi_{j_{2}}\alpha_{j_{2}} 
, ..., \phi_{j_{N_{e}}}\alpha_{j_{N_{e}}})$ and
typically these 1-electron spatial orbitals are expressed in terms of 
atom-centered orbitals $\{|\chi_i >\}$ (themselves often sums or
``contractions'' of gaussian orbitals)
\begin{eqnarray}
|\phi_j  >&=&\sum_i c_{ij}|\chi_i >
\end{eqnarray} 
and the  Hamiltonian matrix elements are expressed in terms of
$\{|\chi_i >\}$.  For a given geometry, the Hartree-Fock orbitals will be a unitary
transformation of any arbitrary starting set of orbitals $\{|\phi_{j}^{0} >\}$
\begin{eqnarray}
|\phi_{k}^{HF}>&=&\sum_j U^{HF}_{jk}|\phi_{j}^{0}\ >\\
|j^{HF}>&=&A(\phi_{j_{1}}^{HF}\alpha_{j_{1}},\phi_{j_{2}}^{HF}\alpha_{j_{2}},...)
\end{eqnarray}
Symbolically, we will write this as $|j^{HF}>=|U^{HF}j^{0}>$
and an arbitrary unitary transformation as $|j>=|U\
j^{0}>$. Since the trace is invariant with respect to unitary
transformations,
\begin{eqnarray}
Q&=&\int\mathbf{dR}^N \sum_j <j^{0}|\exp(-\beta H(\mathbf{R}^{N})|j^{0}>\\
&\propto&\int\mathbf{dR}^N \sum_U \sum_{j^0}<U\ j^0 |\exp(-\beta
H(\mathbf{R}^{N})|U\ j^0 >\label{Usum}
\end{eqnarray}
where the proportionality constant depends on the number of possible
unitary transforms.   Treating the allowed unitary transforms as
rotations, the proportionality constant will then be a proportional to a power of $\pi$ and the sum over $U$ becomes an integral over rotational angles.
As written, it is clear that a
possible algorithm is to view the unitary transforms as just another
degree of freedom to be sampled in a Monte Carlo or molecular dynamics
simulation. 

This expression is exact within the use of a finite set of
states and any related basis set superposition errors.  To make
progress, we will make  approximations to the density matrix elements
$<j|\exp\left[-\beta H(\mathbf{R}^{N})\right]|k>$. The most obvious one is a
Hartree-Fock-like approximation
\begin{eqnarray}
<j|\exp\left[-\beta H(\mathbf{R}^{N})\right]|k>&\approx&\exp\left[-\beta H_{jj}(\mathbf{R}^{N})\right]\delta_{j,k}\label{diag}
\end{eqnarray}
Eqn.~\ref{Usum} then becomes
\begin{eqnarray}
Q_{HF}&\propto&\int\mathbf{dR}^N \sum_U \sum_{j^0 }\exp\left[-\beta
H(\mathbf{R}^{N})_{Uj^0 , Uj^0 }\right]  \label{HF}
\end{eqnarray}
The delta function in Eqn.~\ref{diag} in essence  generates paths  with 0 kinks.  Thus, paths with 0 kinks will correspond to a Hartree-Fock approximation and no electron correlation will be included in a calculation using Eqn.~\ref{HF}.  This expression is an approximation to Eqn.~\ref{Usum} in two
respects; the obvious one being the approximation to the density
matrix and a less apparent one that the sum is not independent of
$U$. In terms of a simulation in which $U$ is sampled, there is an
entropy associated with the different $U$'s which means that not
just $U^{HF}$ will appear, but other $U$'s, which may result in  average
energies that are  higher than  the Hartree-Fock energy.  Of course, sophisticated simulation methods can be used to minimize the effects of this entropy.

To go beyond the Hartree-Fock approximation and include correlation,
consider the discretized version for Q, Eqn.~\ref{originalQ} and
write it in the Born-Oppenheimer approximation as
\begin{eqnarray}
Q\left( P\right)
&=&\int\mathbf{dR}^{N}\left\{\sum_{j}x_{j}^{P}(\mathbf{R}^{N})+
\right. \nonumber \\
&&\left. \sum_{n=2}^{P}\frac{P}{n}\left(
\prod_{i=1}^{n}\sum_{j_{i}}\right) \left(
\prod_{k=1}^{n}t_{j_{k},j_{k+1}}(\mathbf{R}^{N})\right)\right.\nonumber\\
&\times&\left.\sum_{l=1}^{m}\left[ \frac{1}{\left( s_{j_{l}}-1\right) !}\frac{%
d^{s_{j_{l}}-1}}{dx_{j_{l}}^{s_{j_{l}}-1}}\right]
\frac{x_{j_{l}}^{P-1}(\mathbf{R}^{N})}{\prod\limits_{i\neq l}\left(
x_{j_{l}}(\mathbf{R}^{N})-x_{j_{i}}(\mathbf{R}^{N})\right)^{s_{j_{i}}} }\right\} \label{newbeginningQ}
\end{eqnarray}
As a first step, we assume that the most important paths with at least 2 kinks will consist of alternating ground and excited states.  That is, half of the states will be the ground state and the other half will be excited states.
Assuming as previously stated  that
the lowest energy state is non-degenerate, we can write (where now
the summation variable $n$ denotes the number of times the lowest
energy state appears in a path and we suppress the dependence of $x_j
$ and $t_{ij}$ on $\mathbf{R}^{N}$ for notational convenience) 
Eqn.~\ref{newbeginningQ} as
\begin{eqnarray}
Q_2 \left( P\right)  &=&\int\mathbf{dR}^{N}\left\{x_{0}^{P}+\right.  \nonumber \\
&&\left. 2\sum_{n=1}^{P/2}\frac{P}{2n (n-1)!
}\frac{d^{n-1}}{dx_{0}^{n-1}}x_{0}^{P-1}\sum_{n_1 =
0}^{n}\sum_{n_2 = 0}^{n}\cdots \binom{n}{n_1 n_2
\cdots}\prod_{j}\left(\frac{t_{0j}^2 }{(x_0 - x_j )}\right)^{n_j }\right\}\\
&=&\int\mathbf{dR}^{N}\left\{x_{0}^{P}+ \sum_{n=1}^{P/2}\frac{P}{n!
}\frac{d^{n-1}}{dx_{0}^{n-1}}x_{0}^{P-1}\left(\sum_j
\frac{t_{0j}^2 }{x_0 -x_j }\right)^n \right\}\\
&\equiv&\int\mathbf{dR}^{N}\left\{x_{0}^{P}+  \sum_{n=1}^{P/2}\frac{P}{n!
}\frac{d^{n-1}}{dx_{0}^{n-1}}x_{0}^{P-1}\Gamma_{0}^n \right\}\\
&=&\int\mathbf{dR}^{N}\left\{x_{0}^{P}+  \sum_{n=1}^{P/2}\frac{P}{n!
}\sum_{m=0}^{n-1}\binom{n-1}{m}\frac{(P-1)!x_{0}^{P-n+m}}{(P-n+m)!}\frac{d^{m}}{dx_{0}^{m}}\Gamma_{0}^n \right\}
\label{startingpoint}
\end{eqnarray}
where the binomial factor accounts for the number of different
ways to make the different excited states appear in the path and the factor of 2 appears because the first state in the path can be either the ground or an excited state.
Empirically, we have found the the most important term in the sum
over $m$ is the $m=0$ term. This can understood from the
following; the ratio of the $m=1$ term to the $m=0$ is
\begin{eqnarray}
\frac{(n-1)x_0 n\Gamma_1 }{(P-n+1)\Gamma_{0}}
\end{eqnarray}
where
\begin{eqnarray}
\Gamma_1 &\equiv&\frac{d}{dx_0 }\Gamma_0 =-\sum_j \left(\frac{t_{0j}}{x_0 - x_j }\right)^2
\end{eqnarray}
Now for small values of $\beta/P\equiv\epsilon$ we have
\begin{eqnarray}
x_0 &=& <0|\exp(-\epsilon H)|0>\approx 1-\epsilon E_0 \\
t_{oj}&=&<0|\exp(-\epsilon H|0> \approx-\epsilon H_{0j}\\
\Gamma_0 &\approx&\epsilon\sum_j \frac{H_{0j}^2 }{E_0 - E_j } = -\epsilon\Delta E_{MP2}\\
\left|\Gamma_1 /\Gamma_0 \right| &\approx& \frac{1}{\epsilon\Delta E}\\
\frac{(n-1)x_0 n\Gamma_1
}{(P-n+1)\Gamma_{0}}&\approx&\frac{n(n-1)}{\beta\Delta E}
\end{eqnarray}
where $\Delta E_{MP2}$ is the MP2 correction to the Hartree-Fock energy and $\Delta E$ represents a typical difference in energy between
the lowest energy state and one of the excited states appearing in
the sum for $\Gamma_0 $. We typically expect $\beta \Delta E >> 1$
and hence the $m=0$ term to be the most important in
Eqn.~\ref{startingpoint} as long as the sum on n is quickly converging.

Returning to Eqn.~\ref{startingpoint}, we evaluate just the $m=0$
to find the interesting result
\begin{eqnarray}
Q_{2}^{(0)}&\equiv&\int\mathbf{dR}^{N}\left\{x_{0}^{P}+
\sum_{n=1}^{P/2}\binom{P}{n}x_{0}^{P-n}\Gamma_{0}^n \right\}\\
&\approx&\int\mathbf{dR}^{N}\left\{(x_0 + \Gamma_0 )^P \right\}
\end{eqnarray}
where we assume that the sum on $n$ converges sufficiently quickly so
that the sum can be extended from $P/2$ to $P$;  this accuracy of this assumption was checked and confirmed
in the calculations performed in this paper.
Further insight can be gained when $\epsilon$ is very small.  In
this case
\begin{eqnarray}
Q_{2}^{(0)}&\approx&\int\mathbf{dR}^{N}\left\{\left(1-\epsilon\left(E_0 +
\Delta E_{MP2}\right)\right)^P \right\}\\
<E>=-\partial\ln Q/\partial\beta&\approx&<E_0 + \Delta E_{MP2}>_{\mathbf{R}^{N}} =
<E_{MP2}>_{\mathbf{R}^{N}}\label{MP2level}
\end{eqnarray}
where the subscripts on the last averages indicate averaging over the
geometric degrees of freedom.
So paths of alternating ground and excited states, when only the $m=0$ term is
included, should be expected to give an MP2-level result.
Eqn.~\ref{MP2level} will be accurate when a very good
guess to the lowest energy state corresponding to the Hartree-Fock
solution exists for a given geometry.  However, this will not always be
the case during a simulation and
Eqn.~\ref{MP2level} will be refined by including terms beyond the $m=0$
term and more complicated kink patterns in which there is more than a single excited state between occurrences of the ground state.  

We first consider the $m>0$ term in Eqn.~\ref{startingpoint} and
consider more complicated kink patterns later. Note that $m>0$
will include $\Gamma_1 $, the first derivative of $\Gamma_0 $, and
higher order derivatives of $\Gamma_0 $. We expect, and have
verified  in the systems we have studied, that the
major correction to Eqn.~\ref{MP2level}  contains terms with only  $\Gamma_1 $. The derivation of  an expression that includes derivatives up to second order, is possible, but 
not included in herein. Including only those terms in
Eqn.~\ref{startingpoint} of less than second order in derivatives
of $\Gamma_0 $, we find
\begin{eqnarray}
Q_{2}^{(1)}&=&\int\mathbf{dR}^{N}\left\{x_{0}^{P}+  \sum_{n=1}^{P/2}\frac{P}{n!
}\sum_{m=0}^{n-1}\binom{n-1}{m}\frac{(P-1)!x_{0}^{P-n+m}}{(P-n+m)!}\frac{n!
\Gamma_{0}^{n-m}\Gamma_{1}^{m}}{(n-m)!}\right\}\nonumber\\
&=&\int\mathbf{dR}^{N}\left\{x_{0}^P
+\sum_{n=1}^{P/2}\frac{P}{n!}\sum_{k=0}^{n-1}\binom{n-1}{k}\frac{(P-1)!x_{0}^{P-1-k}n!\Gamma_{0}^{k+1}\Gamma_{1}^{n-1-k}}
{(P-1-k)!(k+1)!}\right\}\nonumber\\
&=&\int\mathbf{dR}^{N}\left\{x_{0}^P
+\sum_{k=0}^{P/2-1}\sum_{n=k+1}^{P/2}\frac{P!(n-1)!x_{0}^{P-1-k}\Gamma_{0}^{k+1}\Gamma_{1}^{n-1-k}}
{k!(n-1-k)!(P-1-k)!(k+1)!}\right\}\nonumber\\
&=&\int\mathbf{dR}^{N}\left\{x_{0}^P
+\sum_{k=0}^{P/2-1}\sum_{n=0}^{P/2-k-1}\frac{P!(n+k)!x_{0}^{P-1-k}\Gamma_{0}^{k+1}\Gamma_{1}^{n}}
{k!n!(P-1-k)!(k+1)!}\right\}
\end{eqnarray}
Now
\begin{eqnarray}
\sum_{n=0}^{P/2-k-1}\frac{(n+k)!\Gamma_{1}^{n}} {n!}&=&\frac{d^k
}{d\Gamma_{1}^k }\sum_{n=0}^{P/2-k-1}\Gamma_{1}^{n+k} \approx \frac{d^k
}{d\Gamma_{1}^k }\sum_{n=0}^{\infty}\Gamma_{1}^{n+k}
=\frac{k!}{(1-\Gamma_1 )^{k+1}}
\end{eqnarray}
So we can then find
\begin{eqnarray}
Q_{2}^{(1)}&=&\int\mathbf{dR}^{N}\left\{x_{0}^{P}+\sum_{k=1}^{P/2}\binom{P}{k}x_{0}^{P-k}\left(\frac{\Gamma_0
}{1-\Gamma_1 }\right)^k \right\}
\approx \int\mathbf{dR}^{N}\left\{(x_0 +\frac{\Gamma_0 }{1-\Gamma_1
})^P \right\}
\end{eqnarray}

To include paths with more than one excited state between each
occurrence of the lowest energy state, the above  approach is generalized. We will refer to a portion of the path between 2 occurrences of the ground state as an "excursion".  An excursion will contain one or more excited states.
The development of an expression for the "lowest energy
dominated" (LED) set of paths begins by defining a "weight" associated
with any particular excursion $j$ to be
\begin{eqnarray}
w_j &\equiv&\frac{t_{0a}t_{ab}\cdots t_{z0}}{(x_0 - x_a )(x_0 -
x_b )\cdots (x_0 - x_z )}
\end{eqnarray}
where the excursion $j$ is defined to include the excited states
$a, b, \cdots, z$. Then 
\begin{eqnarray}
Q_{LED}&=&\int\mathbf{dR}^{N}\left\{x_{0}^P +
2\sum_{n=1}^{P/2}\frac{P}{(n-1)!}\frac{d^{n-1}}{dx_{0}^{n-1}}x_{0}^{P-1}\sum_{n_1
}\sum_{n_2 }\cdots \binom{n}{\left\{n_j \right\}}\prod_j
w_{j}^{n_j }\frac{\delta_{n,\sum_j n_j }}{2n+\Delta n(\{n_j \})} \right\}
\nonumber\\ \end{eqnarray}
where $2n+\Delta n(\{n_j \})$ is the number of kinks for a
particular set of excursions.  If the states are well chosen, we
expect the contributions from excursions with greater than one
excited state per excursion to be much less than the contributions
from the one excited state per excursion set of paths.  Thus, as
an \textbf{approximation}, we write
\begin{eqnarray}
\frac{1}{2n+\Delta n(\{n_j \})}&\approx&\frac{1}{2n}
\end{eqnarray}
and we find
\begin{eqnarray}
Q_{LED}&\approx&\int\mathbf{dR}^{N}\left\{x_{0}^P +
\sum_{n=1}^{P/2}\frac{P}{n!}\frac{d^{n-1}}{dx_{0}^{n-1}}x_{0}^{P-1}\Gamma_{0}^{n} \right\}
\end{eqnarray}
where
\begin{eqnarray}
\Gamma_0&=&\sum_j \frac{t_{0j}^{2}}{x_0 - x_j }+\sum_{j\ne
k}\frac{t_{0j}t_{jk}t_{0k}}{(x_0 - x_j )(x_0 -x_k )} + \cdots
\end{eqnarray}
Defining  two matrices, $W_0 $ and $M_0 $
\begin{eqnarray}
(W_{0})_{ij}&=&\frac{t_{0i}t_{0j}}{\sqrt{(x_0 -x_i )(x_0 - x_j
)}}\\
(M_{0})_{ij}&=&\frac{t_{ij}}{\sqrt{(x_0 -x_i )(x_0 - x_j
)}}(1-\delta_{i,j})
\end{eqnarray}
obtains
\begin{eqnarray}
\Gamma_0 &=& Tr\left\{W_0 + W_0 \cdot M_0 + W_0 \cdot M_0 \cdot
M_0 + \cdots \right\}= Tr\left\{W_0 \cdot (I-M_0 )^{-1}\right\}\label{finalQ}
\end{eqnarray}
This sums to infinite order all possible types of excursions from
the lowest energy state, with the proviso that the contributions
from excursions with different numbers of states is a rapidly
decreasing function of the number of states involved in the
excursion. With
\begin{eqnarray}
\Gamma_1 &=& \frac{d}{dx_0 }\Gamma_0 =Tr\left\{W_1 \cdot (I-M_0 )^{-1}+W_0 \cdot (I-M_0 )^{-1} \cdot
M_1 \cdot (I-M_0 )^{-1}\right\}
\end{eqnarray}
we immediately find
\begin{eqnarray}
Q_{LED}&=&\int\mathbf{dR}^{N}\left\{(x_0 + \frac{\Gamma_0 }{1-\Gamma_1
})^P \right\} \propto \int\mathbf{dR}^{N}\sum_{U}\left\{(x_{0}(U) + \frac{\Gamma_{0}(U) }{1-\Gamma_{1}(U)
})^P \right\} \label{led_exact}
\end{eqnarray}
Eqn.~\ref{led_exact} is the principle result of this work and represents an expression that can be used to simulate a molecular system.  There are two computational challenges to using this equation.  The first is that the sum over excited states includes all states and can become a severe bottleneck in a calculation.  This issue can be addressed using a limited set of excited states such as 1- and 2-electron excited states (an approximation made in this work) and by making a slight modification to Eqn.~\ref{led_exact} that will enable excited states to be sampled during the Monte Carlo process.  
$Q_{LED}$ is a function of the ground and excited states, whose total number can be very large. However, it is expected that only a subset of the states will contribute significantly to the partition function and thus we wish to develop a Monte Carlo procedure that will limit the number of excited states used to those with significant contributions to the partition function, as judged by a Monte Carlo simulation.  To do this, first label the excited states in order of decreasing magnitude of $\left[ t_{0j}^2 /(x_0 - x_j )\right] $ (approximately an excited state's contribution to the MP2 energy).  Next the excited states are divided into groups of $N_g $ states, with group 1 corresponding to excited states $1 \dots N_{g}$, group 2 to excited states $N_{g}+1 \dots 2N_{g}$, etc.  If there is a total of $M_{g}$ such groups and  $Q_{LED}(n_{g})$ denotes the result obtained using only the first $n_{g}$ groups of excited states,  $Q_{LED}$ becomes
\begin{eqnarray}
Q_{LED}&=&Q_{LED}(M_g )\nonumber\\
&=&Q_{LED}(M_g )-Q_{LED}(M_g -1)+Q_{LED}(M_g -1)\nonumber\\
&=&\left[Q_{LED}(M_g )-Q_{LED}(M_g -1)\right]+\left[Q_{LED}(M_g -1)-Q_{LED}(M_g -2)\right]+\nonumber \\
&&\cdots+\left[Q_{LED}(1) -Q_{HF}\right]+Q_{HF}\nonumber\\
&\equiv&\Delta Q_{LED}(M_g  ) + \Delta Q_{LED}(M_g -1) +\cdots+\Delta Q_{LED}(1) +\Delta Q_{LED}(0)\nonumber\\
&=&\sum_{j=0}^{M_g } \Delta Q_{LED}(j) \label{kinksum}
\end{eqnarray}
In this notation $\Delta Q_{LED}(0)\equiv Q_{HF}$ and $j$ can be sampled during the Monte Carlo procedure.  If the states are reasonably ordered, the sum over $j$ should converge for a relatively small value of $j$ and the matrices involved in evaluating $\Delta Q_{LED}(j)$ will be manageable in size.
The other time consuming part of any calculation will be the inversion of the matrix $I-M_0 $.  Fortunately, this is amenable to parallel computation using standard linear algebra packages which will aid in the implementation of the method.  
\section{Monte Carlo Sampling Procedure}
\subsection{Rotations of single particle states}
Sampling unitary transformations $U$  was accomplished in the
following way.  Two orbitals $\phi_j $ and  $\phi_i $
were randomly chosen from the list of single particle orbitals. A new
pair of orbitals was formed via a simple unitary transformation
\begin{eqnarray}
\phi^{'}_j &=&\cos\theta\ \phi_j + \sin\theta\ \phi_i \\
\phi^{'}_i &=&-\sin\theta\ \phi_j + \cos\theta\ \phi_i
\end{eqnarray} 
with $\theta$ sampled randomly from $0$ to $2\pi$. These moves were
attempted 40 times each during a Monte Carlo pass (1000 times during the first pass).
\subsection{Addition/removal of kinks}
In this preliminary work only single and double "excitations" from the ground state were considered as candidates for kinks.  That is, the difference between the ground state and an allowed excited state is the transfer of one or two electrons from occupied orbitals to unoccupied orbitals. The following scheme was therefore used for addition and removal of kinks. After sampling the rotations during the first Monte Carlo pass, the ground state was identified (this state did not change during the remainder of the simulation).  A list of states corresponding to double and single excitations was constructed and used for the remainder of the simulation (the list would have been updated if the ground state had changed).  A value of  $N_g = 10$ was used and at each Monte Carlo pass included an attempt to increase or decrease by 1 the value  $j$ in Eqn.~\ref{kinksum}. 
\subsection{Moving atoms}
Each Monte Carlo pass included an attempt to move each atom in turn, as in a standard Monte Carlo simulation.  When an atom move was attempted, the single particle states were no longer orthogonal; the orbitals were orthonormalized using the Gram-Schmidt method.  The step size for each move was 0.03 a.u.
\section{Simulation details}
Monte Carlo simulations were performed using the sampling procedure described above. A temperature of $1/k_{B}T = 3000$ a.u. ($\approx$ 100K) was used; this was high enough to allow the relatively large geometric changes required to find the global minimum geometries, but not so high as to introduce large vibrational motion.  All molecules were started in planar (and linear in the case of H$_{2}$O) geometries to test the ability of the method to find the correct geometry. $P = 3 \times 10^{10}$ and 1000 Monte Carlo passes were performed and averages were computed using the last 500 passes.  One and two-electron integrals were evaluated using the C version programs included in PyQuante \cite{pyquante} and SCALAPACK\cite{scalapack}  routines were used to perform matrix inversions.  The 6-31G basis set was used and two simulations were performed for each molecule.  In the first, no kinks were allowed resulting in a simulation using Eqn.~\ref{HF}.  A second simulation was performed using Eqn.~\ref{kinksum} providing a simulation that included correlation.  Calculations used 16 processors on the SuperHelix computer at LSU (www.cct.lsu.edu.)  Correlation lengths of the energy and bond lengths ranged from 50-150 Monte Carlo passes, reasonable values given the Monte Carlo step size.

\section{Results and Discussion}
 
\begin{figure}[htbp]
\begin{center}
\includegraphics*[scale=0.5,keepaspectratio=true]{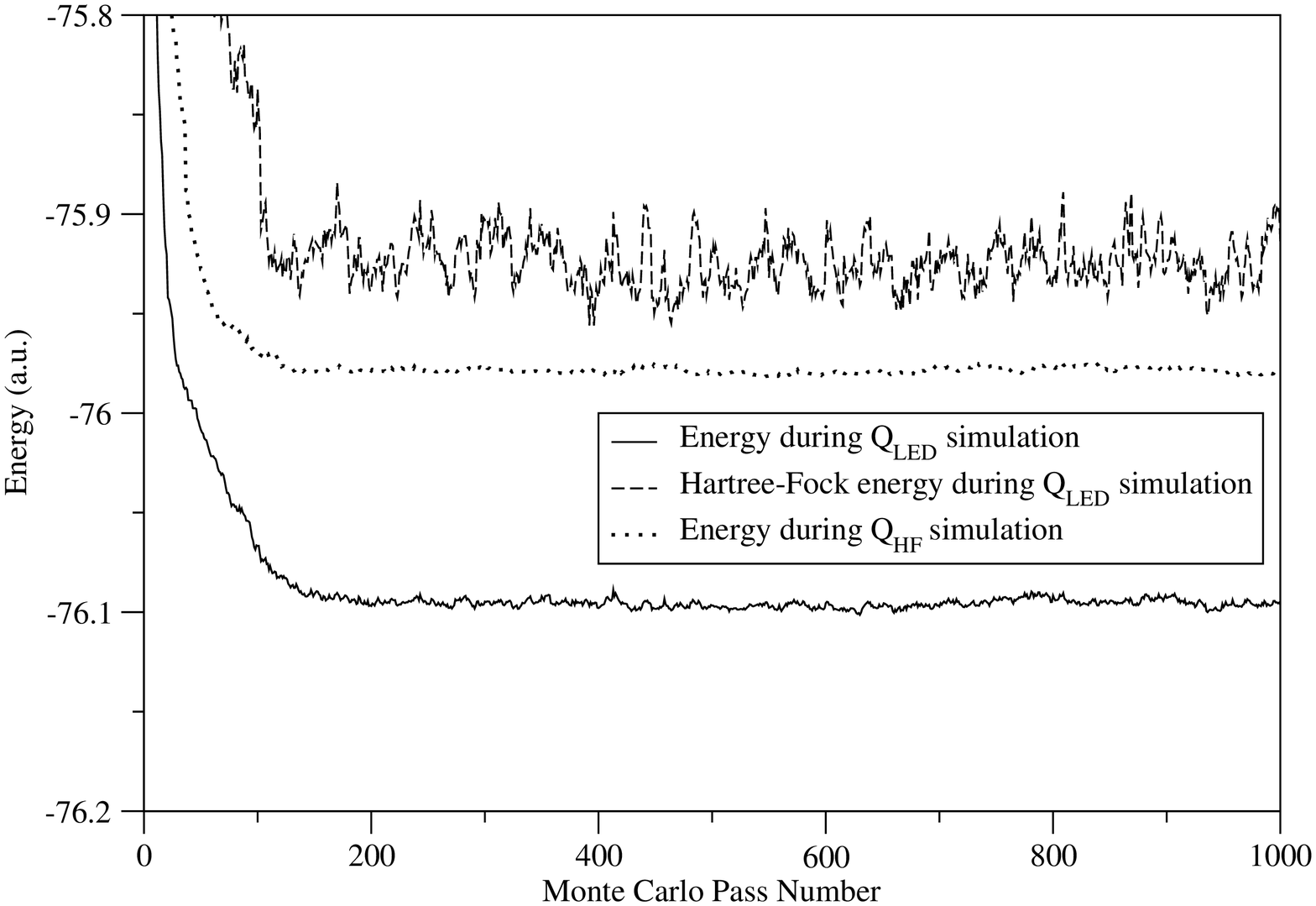}
\caption{\label{h2o_fig}}
\end{center}
\end{figure}

\begin{figure}[htbp]
\begin{center}
\includegraphics*[scale=0.5,keepaspectratio=true]{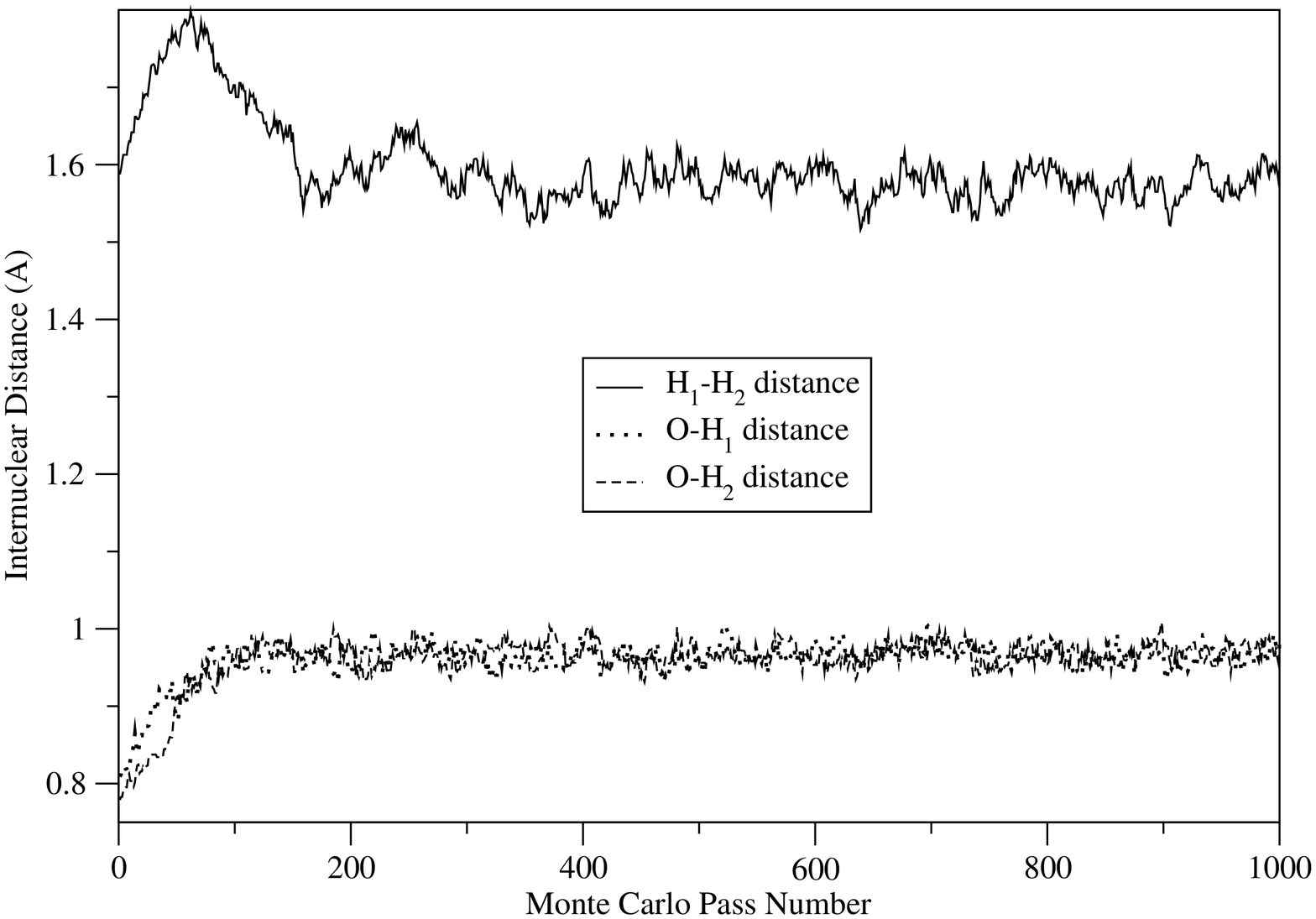}
\caption{\label{h2obond_fig}}
\end{center}
\end{figure}

 The first molecule studied was H$_2 $O. Started in a linear geometry, the molecule quickly became bent and adopted the expected geometry. Fig.~\ref{h2o_fig} shows the variation of total energy and ground state (HF) energy during the simulation using Eqn.~\ref{kinksum}.  Fig.~\ref{h2obond_fig} displays the evolution of the different internuclear distances during the simulation.  Several important features are evident from these figures.  First, the electronic and geometric degrees of freedom evolve to their equilibrium values in a similar number of Monte Carlo passes.  Second, the fluctuations in the Hartree-Fock energy are quite large ($\approx$.03 a.u. $\approx$ 19 kcal/mol), demonstrating that the Monte Carlo procedure does not require a particularly accurate estimate for  the Hartree-Fock ground state.  Third, despite the fluctuations in the Hartree-Fock energy, the correlated energy has small fluctuations, which is to be expected of a good algorithm.  From Fig.~\ref{h2o_fig} and Table~\ref{h2o_table} we can compare the correlated and uncorrelated methods used in this study.  The Hartree-Fock estimator (Eqn.~\ref{HF}) results in  fluctuations in the energy estimator that are small and comparable to the fluctuations in the total energy estimator using Eqn.~\ref{kinksum}.  
 
 Tables~\ref{h2o_table}-\ref{ch4_table} summarize the energies and geometries for all simulations.  For comparison, we have calculated the 0 K energies (with and without zero point energy correction) and geometries using Gaussian 98\cite{g98}.    
 
 Comparing the energies calculated using the path integral simulations with those obtained using Gaussian 98, we find that the path integral results using $Q_{LED}$ are in close agreement with MP2 level energies and in much better agreement with MP2 energies than are the DFT results.  The path integral energies lie above the 0 K \textit{ab initio} energies, as expected due to the entropy associated with the unitary transformations and finite temperature (classical) vibrational effects.  The energies are below the zero point corrected \textit{ab initio} energies due to the large zero point energy corrections in these molecules.  The average geometries are in good agreement with \textit{ab initio}  results both in absolute bond lengths and angles and in the differences between Hartree-Fock and correlated (MP2/DFT) results, particularly in light of the \textit{ab initio} bond lengths and angles being appropriate at 0 K and the path integral results appropriate at 100 K.  The largest difference between the path integral and \textit{ab initio} geometries occurs in NH$_{3}$, where the bond angles are larger by 2 - 3 degrees in the path integral calculations.  Since the \textit{ab initio} geometries are obtained at 0 K and there is a relatively low frequency vibrational mode in NH$_{3}$, we performed a path integral calculation at a lower temperature to see if the average angles from a simulation came into better agreement with the 0 K results.  A simulation at $\beta = 10000\  a.u.$ ($\approx$ 30 K) resulted in energies and geometries shown in Table~\ref{nh3_table}.  The bond angles are found to be in much better agreement with the 0 K values.  In addition, the energies are in better agreement with \textit{ab initio} results.
 
The Monte Carlo procedure selects excited states that make a significant contribution to the partition function.  The number of possible 1- and 2- electron excited states range from 2240 for H$_2 $O to 5040 for CH$_4 $.  The average number of excited states  in the simulation ranges from 400-500.  Thus, the Monte Carlo procedure is able to restrict the computational effort and bodes well for the scaling in larger systems.
 
\section{Conclusion}
The kink-based path integral formulation has been extended to molecular systems.  An approximate infinite order summation is used to include Hartree-Fock-like excited states in the ground state, correlated wavefunction.  This procedure is necessary because all matrix elements are referenced to atom based primitive orbitals, which makes storage of the full N-electron density matrix too time consuming to be feasible.  The estimator developed using this approach was compared to a Hartree-Fock-like method.  In terms of geometries, the  correlated method compares well with standard \textit{ab initio} MP2 results (and are significantly better than DFT level results) and the Hartree-Fock-like geometries are in good agreement with 0 temperature \textit{ab initio} Hartree-Fock calculations.  
The treatment of geometric and electronic degrees of freedom on the same footing is a strength of this method.  These initial results suggest this approach, combined with parallel computing, will provide an important alternative to standard \textit{ab initio} methods, as well as the very successful Car-Parrinello DFT method.  

A direct comparison between the computational effort of conventional \textit{ab initio} approaches and the kink-based method is somewhat difficult since the bottleneck in a conventional simulation is the solution of the Hartree-Fock problem while in the path integral calculation a matrix inversion is the time consuming part of the calculation.  It is possible, though, to discuss possible improvements to the kink-based approach.  The time to invert a matrix scales with the third power of the number of excited states, which (even with parallel computing) may prove to be a bottleneck to computational efficiency.  However importance sampling using an importance function such as 
\begin{eqnarray}
Q_{LED}^{app}&=&\int\mathbf{dR}^{N}\left\{(x_0 + \frac{\Gamma_{0}^{app} }{1-\Gamma_{1}^{app}
})^P \right\}\\
\Gamma_{0}^{app}&=&Tr\left\{W_0 \cdot (I+M_0 )\right\}\\
\Gamma_{1}^{app}&=&\frac{d}{dx_0 }\Gamma_{0}^{app}\label{led_app}
\end{eqnarray}
\begin{eqnarray}
Q_{LED}&=&\int\mathbf{dR}^{N}\left\{\frac{(x_0 + \frac{\Gamma_0 }{1-\Gamma_1
})^P }{(x_0 + \frac{\Gamma_{0}^{app} }{1-\Gamma_{1}^{app}
})^P}\right\}(x_0 + \frac{\Gamma_{0}^{app} }{1-\Gamma_{1}^{app}
})^P
\label{led_imp}\end{eqnarray}
reduces the computational effort significantly because only matrix multiplications are involved in the actual Monte Carlo moves.  Some initial studies with this importance function showed a significant decrease in computational effort with only a minor decrease in precision.

Also of interest is the question of size consistency.  If a system is duplicated $n$ times into $n$ non-interacting systems, the partition function becomes the product of the individual partition functions, which will guarantee size consistency.  Also of interest from a size consistency point of view is whether the Monte Carlo estimator reaches this factorization limit.  An examination of the leading terms in Eqn.~\ref{startingpoint} indicates that $\Gamma_0 $ scales with $n$ and $\Gamma_1 $ is independent of $n$.  The latter feature can be understood in the following way.  The denominators in $\Gamma_0 $ do not scale with $n$ because the allowed excited states $j$ are localized on one of the $n$ systems.  However, the derivative necessary to obtain $\Gamma_1 $ scales with $1/n$.  Since the number of excited states also scales with $n$, $\Gamma_1 $ will be independent of $n$.  Therefore, the Monte Carlo estimator for $n$ systems becomes
\begin{eqnarray}
\left(x_{0}^n + \frac{n \Gamma_ 0}{1-\Gamma_1 }\right)^P &\approx&x_{0}^{nP}\exp\left(\frac{P n \Gamma_0 }{1-\Gamma_1 }\right)
\end{eqnarray}
This clearly is size consistent.

\noindent\acknowledgements{It is a pleasure to acknowledge Professor Neil Kester for useful discussions. The Gaussian 98 calculations were performed by Cheri McFerrin. This work was partially supported by NSF grant CHE 9977124 and by the Center for Computation and Technology at LSU.}
\newpage
\section{References}
\bibliography{rwh}

\clearpage
\begin{table}
\caption{Energies, average number of excited states included in the path integral calculation ($<N_{s}>$), and structural parameters for H$_2 $O.  All energies and distances are in atomic units and numbers in parenthesis represent 2 standard deviations (95\% confidence limits).  $<$E$>$ is the energy, including correlation, $<$E$_{HF}>$ is the energy of the lowest energy state, $<d_{HH}>$ is the average H-H bond length,  $<d_{OH}>$ is the average O-H bond length, and $<\alpha_{HOH}>$ is the average H-O-H angle.  Path integral calculations were performed at $\beta = 3000$  a.u. ( $\approx$  T = 100 K).  \textit{Ab initio} results were obtained using Gaussian 98\cite{g98} and are given with and without the zero point energy correction (zpe).}
\label{h2o_table}
\begin{tabular}{|l||l|l|l|l|l|l|}
\hline
H$_{2}$O&$<$E$>$&$<$E$_{HF}>$&$<N_{s}>$&$<d_{HH}>$&$<d_{OH}>$&$<\alpha_{HOH}>$\\ \hline \hline
PI, $Q_{HF}$ (Eqn.~\ref{HF})&&-75.979(1)&0&1.57(1)&0.951(4)&111(2)\\ \hline
PI, $Q_{LED}$  (Eqn.~\ref{kinksum})&-76.096(2)&-75.93(1)&578&1.57(1)&0.968(4)&109(1)\\  \hline
\textit{ab initio} HF\ \  (with zpe) &&-75.963&&1.57&0.95&112\\ 
\ \ \ \ \ \ \ \ \ \ \ \ \ \ \ \ \ \ (without zpe) && -75.985 &&&&\\ \hline
\textit{ab initio} DFT (B3LYP, with zpe)&-76.366&&&1.58&0.98&108\\ 
\ \ \ \ \ \ \ \ \ \ \ \ \ \ \ \ \ \ \ (without zpe) &-76.386&  &&&&\\ \hline
\textit{ab initio} MP2 (with zpe)&-76.092&&&1.59&0.97&109\\ 
\ \ \ \ \ \ \ \ \ \ \ \ \ \ \ \ \ \ \ (without zpe) &-76.113&  &&&&\\ \hline
\hline
\end{tabular}
\end{table}
\clearpage
\begin{table}
\caption{Energies, average number of excited states included in the path integral calculation ($<N_{s}>$), and structural parameters for NH$_3 $.  All energies and distances are in atomic units and numbers in parenthesis represent 2 standard deviations (95\% confidence limits).  $<$E$>$ is the energy, including correlation, $<$E$_{HF}>$ is the energy of the lowest energy state, $<d_{HH}>$ is the average H-H bond length,  $<d_{NH}>$ is the average N-H bond length, and $<\alpha_{HNH}>$ is the average H-N-H angle.   Path integral calculations were performed at $\beta = 3000$  a.u. ( $\approx$  T = 100 K) except as noted.  \textit{Ab initio} results were obtained using Gaussian 98\cite{g98}  and are given with and without the zero point energy correction (zpe).}
\label{nh3_table}
\begin{tabular}{|l||l|l|l|l|l|l|}
\hline
NH$_{3}$&$<$E$>$&$<$E$_{HF}>$&$<N_{s}>$&$<d_{HH}>$&$<d_{NH}>$&$<\alpha_{HNH}>$\\ \hline \hline
PI, $Q_{HF}$   (Eqn.~\ref{HF})&&-56.156(1)&0&1.71(1)&0.989(4)&119(1)\\ \hline
PI, $Q_{LED}$  (Eqn.~\ref{kinksum})&-56.240(2)&-56.09(1)&417&1.70(1)&1.000(3)&117(1)\\ \hline
PI, $\beta=10000\ a.u., Q_{LED}$  (Eqn.~\ref{kinksum})&-56.2760(2)&-56.140(2)&951&1.694(2)&1.009(1)&114.2(2)\\ \hline
\textit{ab initio} HF\ \  (with zpe) &&-56.129&&1.68&0.99&116\\
\ \ \ \ \ \ \ \ \ \ \ \ \ \ \ \ \ \ (without zpe) && -56.166 &&&&\\ \hline
\textit{ab initio} DFT (B3LYP, with zpe)&-56.498&&&1.71&1.01&116\\ 
\ \ \ \ \ \ \ \ \ \ \ \ \ \ \ \ \ \ \ (without zpe) &-56.532&  &&&&\\ \hline
\textit{ab initio} MP2 (with zpe)&-56.244&&&1.70&1.01&114\\ 
\ \ \ \ \ \ \ \ \ \ \ \ \ \ \ \ \ \ \ (without zpe) &-56.280&  &&&&\\ \hline
\hline
\end{tabular}
\end{table}
\clearpage
\begin{table}
\caption{Energies, average number of excited states included in the path integral calculation ($<N_{s}>$), and structural parameters for  CH$_4 $.  All energies and distances are in atomic units and numbers in parenthesis represent 2 standard deviations (95\% confidence limits).  $<$E$>$ is the energy, including correlation, $<$E$_{HF}>$ is the energy of the lowest energy state, $<d_{HH}>$ is the average H-H bond length,  $<d_{CH}>$ is the average C-H bond length, and $<\alpha_{HCH}>$ is the average H-C-H angle.   Path integral calculations were performed at $\beta = 3000$  a.u. ( $\approx$  T = 100 K)  \textit{Ab initio} results were obtained using Gaussian 98\cite{g98}  and are given with and without the zero point energy correction (zpe).}
\label{ch4_table}
\begin{tabular}{|l||l|l|l|l|l|l|}
\hline
CH$_{4}$&$<$E$>$&$<$E$_{HF}>$&$<N_{s}>$&$<d_{HH}>$&$<d_{CH}>$&$<\alpha_{HCH}>$\\ \hline \hline
PI, $Q_{HF}$   (Eqn.~\ref{kinksum})&&-40.168(1)&0&1.766(5)&1.082(4)&109.4(1)\\ \hline
PI, $Q_{LED}$  (Eqn.~\ref{kinksum})&-40.254(4)&-40.11(1)&525&1.782(6)&1.092(4)&109.4(4)\\  \hline
\textit{ab initio} HF (with zpe)&&-40.132&&1.78&1.08&110\\ 
\ \ \ \ \ \ \ \ \ \ \ \ \ \ \ \ \ (without zpe) && -40.181 &&&&\\ \hline 
\textit{ab initio} DFT (B3LYP, with zpe)&-40.465&&&1.79&1.09&109\\ 
\ \ \ \ \ \ \ \ \ \ \ \ \ \ \ \ \ \ \ \ (without zpe) &-40.511&  &&&&\\ \hline
\textit{ab initio} MP2 (with zpe)&-40.233&&&1.79&1.10&109\\ 
\ \ \ \ \ \ \ \ \ \ \ \ \ \ \ \ \ \ \ (without zpe) &-40.279&  &&&&\\ \hline
\hline\end{tabular}
\end{table}

\clearpage
\begin{center}
Figure Captions
\end{center}
Figure 1.  Energies during different Monte Carlo simulations of H$_2 $O. Energy during Q$_{LED}$ simulation is the energy during a simulation using Eqn.~\ref{kinksum}, Hartree-Fock energy during Q$_{LED}$ simulation is the energy of the lowest energy state during a simulation using Eqn.~\ref{kinksum}, and energy during Q$_{HF}$ simulation is the energy during a simulation using Eqn.~\ref{HF}.\\
Figure 2.  Internuclear distances during different Monte Carlo simulation of H$_2 $O using Q$_{LED}$.  The two hydrogen atoms are labeled H$_1 $ and H$_2 $.  The initial values of the interatomic distances correspond to the initial linear geometry.
\clearpage

\clearpage

\end{document}